\documentclass[twocolumn]{aastex6}
\usepackage{gensymb}
\usepackage{natbib}
\begin{document}
\submitted{}
\title{A new spectral feature on the trailing hemisphere of Europa at 3.78 $\micron$}

\author{Samantha K. Trumbo, Michael E. Brown, and Patrick D. Fischer }
\affil{Division of Geological and Planetary Sciences, California Institute of Technology, Pasadena, CA 91125, USA}
\author{Kevin P. Hand}
\affil{Jet Propulsion Laboratory, California Institute of Technology, Pasadena, CA 91109, USA}

\begin{abstract}

We present hemispherically resolved spectra of the surface of Europa from $\sim$3.1--4.13 $\micron$, which we obtained using the near infrared spectrometer NIRSPEC on the Keck II telescope. These include the first high-quality L-band spectra of the surface to extend beyond 4 $\micron$. In our data we identify a previously unseen spectral feature at 3.78 $\micron$ on the trailing hemisphere. The longitudinal distribution of the feature is consistent with that of a radiolytic product created by electron or Iogenic ion bombardment. This feature is coincident with an absorption feature of SO$_2$ frost seen in both laboratory spectra and spectra of Io. However, the corresponding, typically stronger 4.07 $\micron$ feature of SO$_2$ frost is absent from our data. This result is contrary to the suggested detection of SO$_2$ at 4.05 $\micron$ in Galileo NIMS data of the trailing hemisphere, which was severely affected by radiation noise. We use simple spectral modeling to argue that the 3.78 $\micron$ feature is not easily explained by the presence of SO$_2$ frost on the surface. We explore alternative explanations and discuss other potential candidate species.

\end{abstract}

\keywords{planets and satellites: composition --- planets and satellites: individual (Europa) --- planets and satellites: surfaces}

\section{Introduction}\label{sec:intro}
The surface composition of Europa is of prime interest, because it may ultimately constrain the composition of the ocean below. However, because Europa is located within Jupiter's magnetosphere, its surface is continuously bombarded with energetic charged particles trapped within Jupiter's rapidly rotating magnetic field. These include both high-energy electrons and lower-energy sulfur ions originating from the volcanos of Io \citep{Paranicas2001, Paranicas2002}. The resultant interactions drive much of Europa's known surface chemistry by radiolytically processing the surface and creating several new products \citep[e.g.][]{Carlson2002, Carlson2005, Paranicas2009}. Europa is tidally locked to Jupiter, and Jupiter rotates much faster than Europa completes its orbit (11.2 hours synodic to Europa vs. Europa's 3.55 day orbital period). Thus, most particles trapped within Jupiter's co-rotating magnetosphere preferentially impact the trailing hemisphere, producing the ``bullseye" pattern of radiolytically produced hydrated material observed by Galileo NIMS (Near Infrared Mapping Spectrometer) \citep[e.g.][]{Paranicas2001, Paranicas2009, Carlson2009}. Laboratory investigations of radiolytic chemistry in sulfur-water ice mixtures suggest hydrated sulfuric acid as the dominant product, and indeed this species fits the NIMS spectra well \citep{Carlson1999, Carlson2002, Carlson2005}.

To date, much of the compositional information, including that indicative of radiolytic chemistry, has been deduced from NIMS observations \citep[e.g.][]{McCord1998, McCord1999, Carlson2002, Carlson2005, Carlson2009, HansenMcCord2008}. In addition to the hypothesized trailing hemisphere sulfuric acid \citep{Carlson1999}, other likely detections include CO$_2$ at 4.25 $\micron$ on both the leading and trailing hemispheres \citep{McCord1998, Smythe1998, Hand2007, HansenMcCord2008} and H$_2$O$_2$ at 3.5 $\micron$ on the leading hemisphere \citep{Carlson1999Perox}. However, observations in the 3--5 $\micron$ range of the trailing hemisphere were severely limited by the intense radiation environment at Europa's orbit, and, as a result, the data are of low quality. Recent ground-based observations have provided the best-quality data in this wavelength region. \citet{HandBrown2013} presented the first high quality 3--4 $\micron$ spectra of Europa's surface, in which H$_2$O$_2$ was hemispherically resolved across four nights of observation. These data also revealed a previously unseen feature at 3.78 $\micron$, which is the focus of this paper. 

\floattable
\begin{deluxetable*}{cccccc}
\tablecaption{Table of Observations\label{table:obs}}
\tablecolumns{6}
\tablenum{1}
\tablewidth{0pt}
\tablehead{
\colhead{Date} & \colhead{Target} & \colhead{Time} & \colhead{Airmass} & \colhead{Longitude} & \colhead{Int. Time}\\
\colhead{(UT)} & \colhead{ } & \colhead{Start/End} & \colhead{Start/End} & \colhead{Range} & \colhead{(s)}}

\startdata
2011 Sep 17 & Europa & 12:45/12:54 &  1.01/1.01 & 315 - 316 & 840\\
 & HD 9866 & 13:56 & 1.10 & & 40\\
2011 Sep 18 & Europa & 11:47/13:58 & 1.06/1.02 & 52 - 62 & 6240\\
 & HD 9866 & 11:40 & 1.02 & & 40\\
2011 Sep 19 & Europa & 12:50/13:20 & 1.01/1.01 & 158 - 161 & 1440\\
 & HD 9866 & 11:25 & 1.02 &  & 40\\
2011 Sep 20 & Europa & 11:55/13:57 & 1.04/1.04 & 256 - 264 & 5760\\
 & HD 9866 & 11:48 & 1.04 & & 40\\
2013 Nov 24 & Europa & 13:03 /15:30 & 1.00/1.09 & 245 - 255 & 3800\\
 & G91-3 & 13:20/13:34 & 1.00/1.11 & & 750\\
\enddata
\end{deluxetable*}

The feature is localized to the trailing hemisphere, which suggests it might be a radiolytic product produced via bombardment by electrons or Iogenic ions. However, whether the new feature is actually indicative of a previously undetected  species or is simply a result of a component already confirmed at other wavelengths, is not immediately clear. Perhaps the most obvious candidate is SO$_2$ frost, which has a 3.78 $\micron$ absorption \citep[e.g.][]{NashBetts1995} and has been confirmed at UV wavelengths with a trailing hemisphere enhancement \citep{Lane1981,Noll1995,Hendrix1998}. SO$_2$ dominates the 3--4 $\micron$ spectrum of Io \citep[e.g.][]{Howell1989, NashBetts1995, Carlson1997Io} and is an expected product of Iogenic sulfur implantation on Europa \citep{Carlson2002, Carlson2005}. 

The spectrum of SO$_2$ frost has a much stronger feature at $\sim$4.07 $\micron$, which is beyond the data of \citet{HandBrown2013}. \citet{HansenMcCord2008} report a marginal detection of SO$_2$ at 4.05 $\micron$ in NIMS spectra of the trailing hemisphere. These data were taken during distant Europa passes to minimize radiation noise, but they are still of much lower quality than can be achieved from the ground. Furthermore, the reported SO$_2$ band strengths of up to 40\% are inconsistent with the 0.1\% average 4.07 $\micron$ band strength predicted from the UV detections \citep{Sack1992,Carlson2009}. The only reported ground-based L-band detection of SO$_2$ on Europa \citep{Carter2013} is a result of erroneous observations of the surface of Io and should be disregarded. 

To investigate the possibility that the 3.78 $\micron$ feature is due to SO$_2$, we obtained high-quality L-band spectra of the trailing hemisphere from $\sim$3.33--4.13 $\micron$ using the near infrared spectrograph NIRSPEC at the W. M. Keck Observatory \citep{McLean1998}. In contrast with the results of \citet{HansenMcCord2008}, the 4.07 $\micron$ feature of SO$_2$ is not visible in our spectrum. We apply simple spectral models to argue that the 3.78 $\micron$ feature is not explained by SO$_2$ and, thus, is highly indicative of an unidentified radiolytic surface constituent. We discuss potential candidates in Section \ref{sec:alternatives}.

\section{Observations and Data Reduction}\label{sec:methods}
We observed the trailing hemisphere of Europa on 2013 Nov. 24, using the near-infrared spectrograph NIRSPEC on the KECK II telescope. We used the 42" x 0.57"  slit in low-resolution mode (R $\sim$ 2000), covering a wavelength range of $\sim$3.3--4.15 $\micron$ in the L-band. At the time of observation, Europa had an angular diameter of 0.97", such that roughly 60$\%$ of the disk fit within the slit. For telluric calibration,  we observed G91-3, a V = 7.4 G2V star that was $\sim$3.6$\degr$ away from Europa on the sky. We observed both targets in an ABBA nodding pattern. Each Europa pointing consisted of 100 half-second coadds, and each calibrator pointing consisted of  50 half-second coadds. We also present previously published L-band spectra from \citet{HandBrown2013}. Details of the corresponding observations and analyses can be found within the referenced paper, although  a summary of all observations is presented in Table \ref{table:obs}.

We performed all new analysis using custom Python codes, following the standard procedures of image rectification, image pair subtraction, residual sky subtraction, and telluric and wavelength calibration. We used the Astropy \citep{Astropy} and skimage.transform \citep{skimage} packages for image display and rectification, respectively. For wavelength calibration, we used an ATRAN atmospheric transmission spectrum \citep{ATRAN}. We leave off a small region of data at the long-wavelength end of our spectrum due to excessive thermal background levels. We averaged the spectra over the entire longitude range observed to maximize signal-to-noise.
\begin{figure*}
\figurenum{1}
\plotone{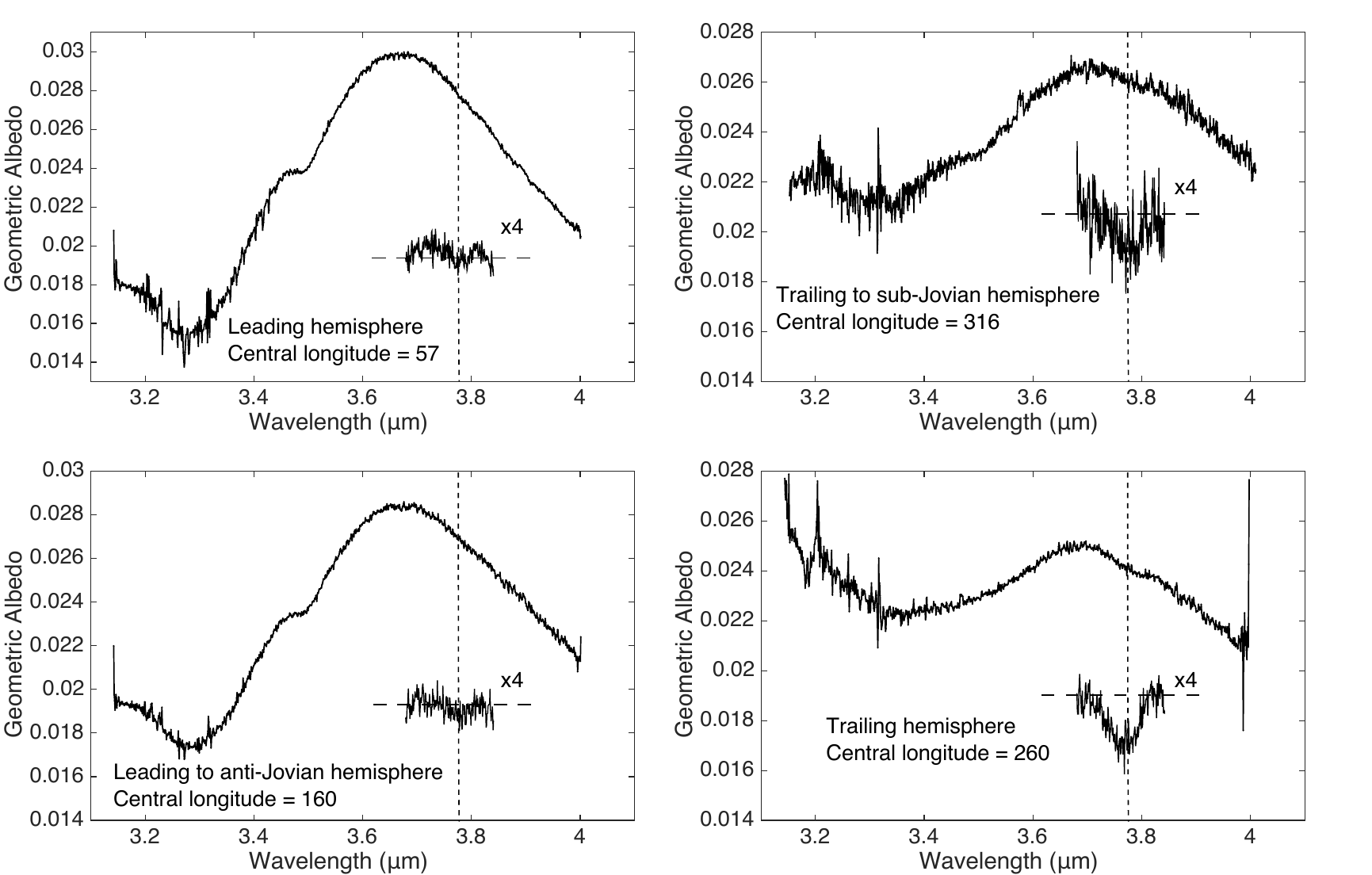}
\caption{Averaged hemispherically resolved spectra of \citet{HandBrown2013}. Data are normalized to the geometric albedo of the leading hemisphere peak, as measured by NIMS. The central longitude of each spectrum is included. Dotted lines mark the 3.78 $\micron$ position. The 3.78 $\micron$ feature is most prominent in the trailing hemisphere spectrum. \label{fig:spectra}}
\end{figure*}

\section{Spectral Modeling: SO$_2$}\label{sec:so2}
\begin{figure}
\figurenum{2}
\plotone{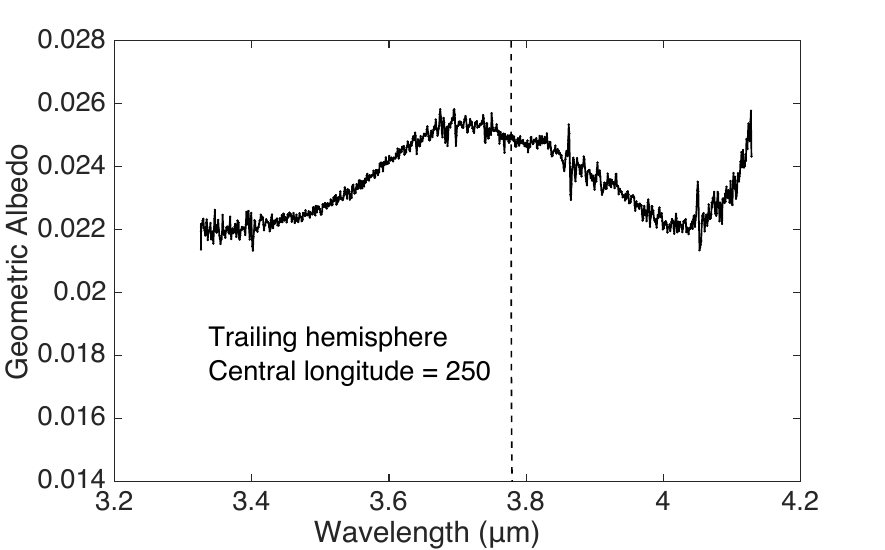}
\caption{Averaged trailing hemisphere spectrum out to 4.13 $\micron$. The data are scaled such that the peak matches the geometric albedo observed by NIMS. The central longitude of the observations is also included. The 3.78 $\micron$ feature is clearly visible (indicated by dashed line), but there is no obvious absorption at either 3.56 or 4.07 $\micron$. \label{fig:spectrum}}
\end{figure}

The hemispherically resolved spectra originally presented in \citet{HandBrown2013} are reproduced in Figure \ref{fig:spectra}. The 3.78 $\micron$ feature is clearly visible in the trailing hemisphere spectrum centered at 260$\degr$ west longitude and is just visible in the trailing to sub-Jovian spectrum centered at 316$\degr$ west longitude. However, it is very weak or absent from the leading and leading to anti-Jovian spectra. Figure \ref{fig:spectrum} shows our average spectrum of the trailing hemisphere out to 4.13 $\micron$. The data are scaled such that the maximum value is roughly the geometric albedo of the trailing hemisphere peak, as measured by NIMS. The 3.78 $\micron$ feature is again visible, but the 4 $\micron$ region is smooth and absent of any visible SO$_2$ absorptions. As these data are somewhat noisier than the 2011 data of \citet{HandBrown2013}, we use the latter for all subsequent analyses of the 3.78 $\micron$ absorption.

\begin{figure}
\figurenum{3}
\plotone{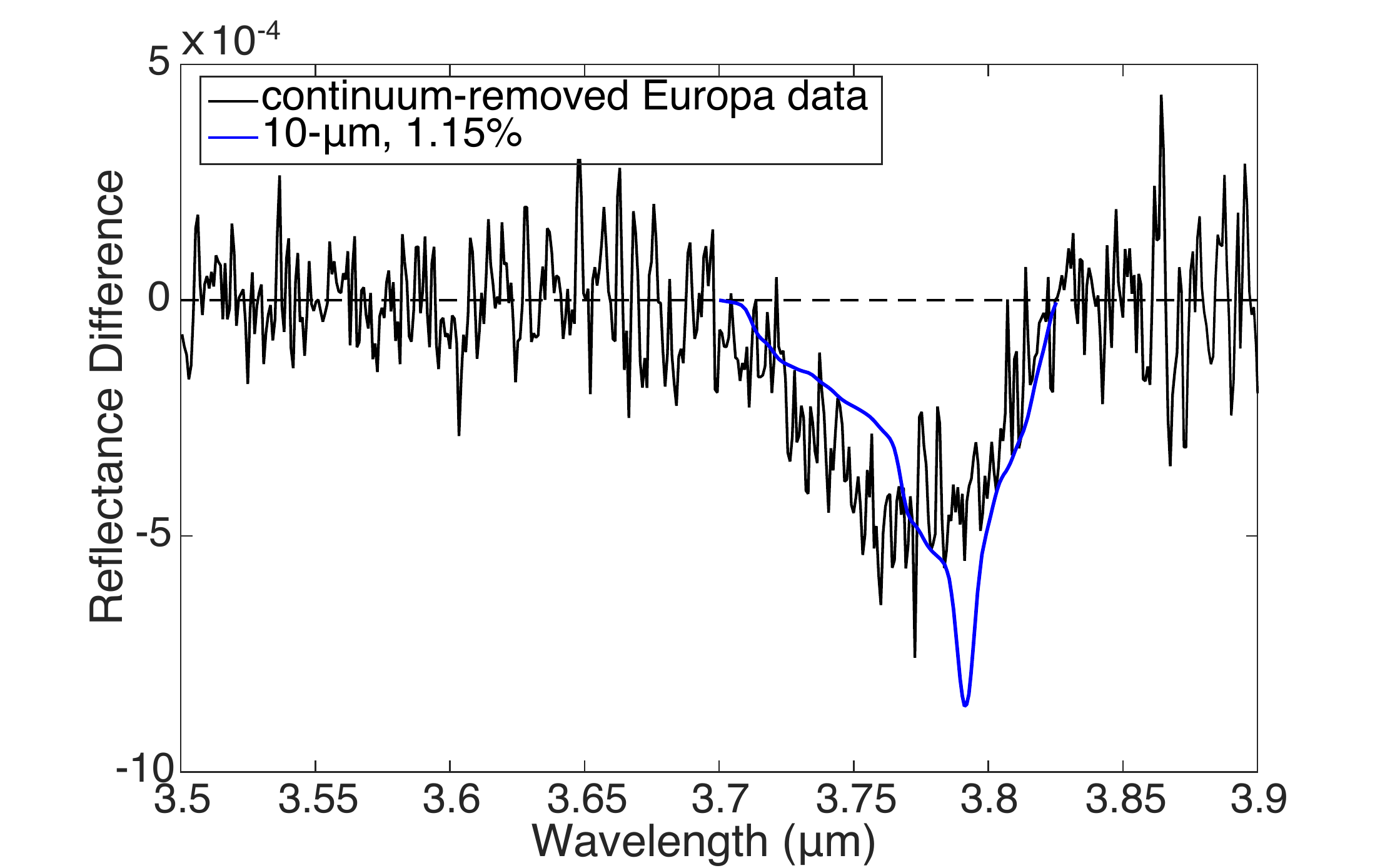}
\caption{The 3.78 $\micron$ feature on the trailing hemisphere of Europa after continuum removal. Overlaid is the SO$_2$ 3.78 $\micron$ feature corresponding to an effective grain size of 10 $\micron$. It has been scaled to match the band area of our observed absorption feature. The scaling factor is presented in the legend. Continuum removal was performed on the 2011 data of \citet{HandBrown2013}, because it has the highest signal-to-noise. \label{fig:378fits}}
\end{figure}

\begin{figure}
\figurenum{4}
\plotone{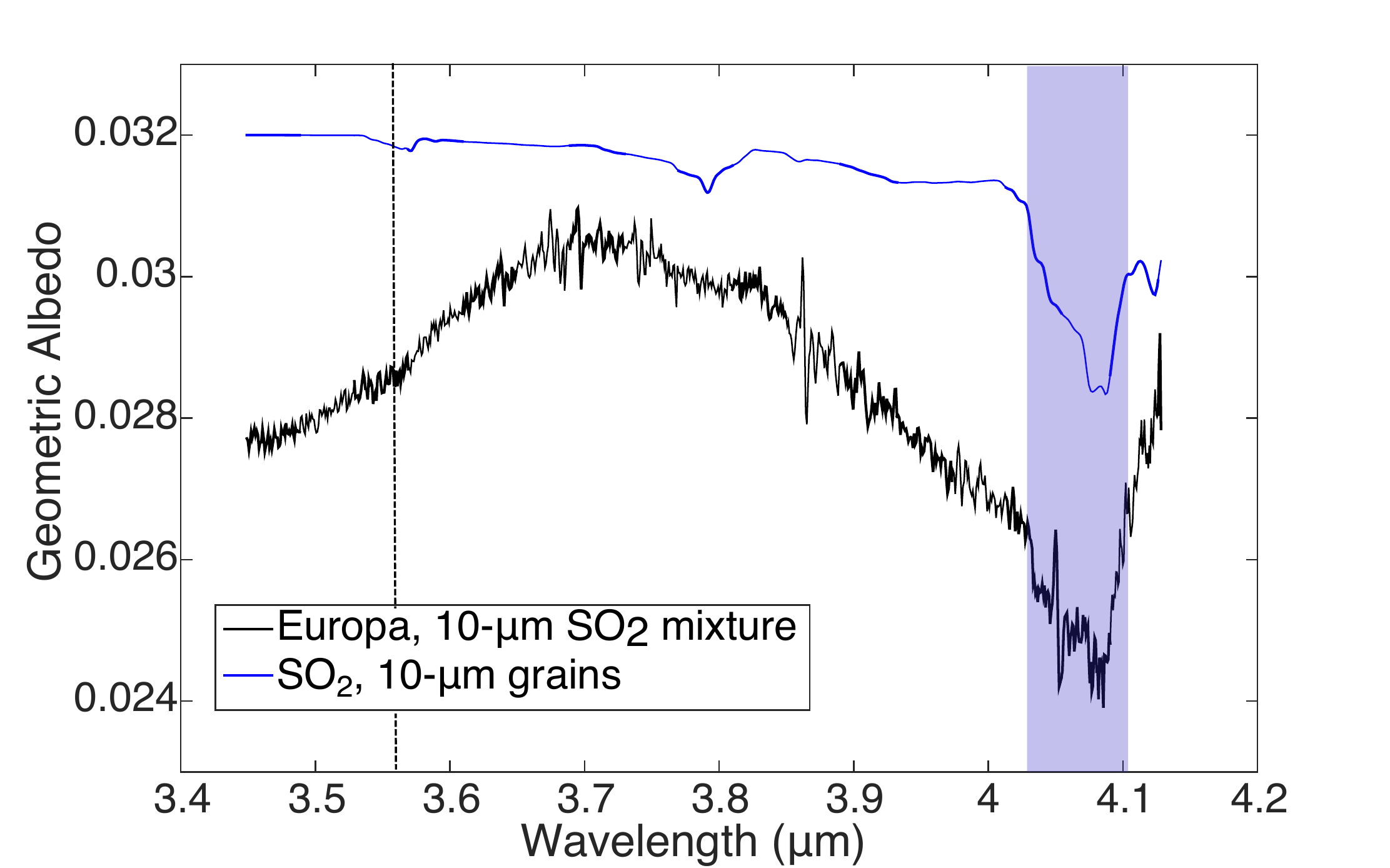}
\caption{Linear mixture of the 2013 trailing hemisphere Europa spectrum with SO$_2$ at a 10 $\micron$ effective grain size is shown in black. The mixing ratio reflects the amount needed to produce the observed 3.78 $\micron$ band area. For comparison, a simulated SO$_2$ spectrum for a 10 $\micron$ effective particle size is shown in blue. It has been scaled by the same mixing ratio and arbitrarily shifted along the y-axis. The 4.07 $\micron$ feature of SO$_2$, highlighted in blue, is clearly distinguishable from the continuum shape of the mixture. The location of the 3.56 $\micron$ SO$_2$ feature is indicated by the dashed line. We argue that SO$_2$ grain sizes less than or equal to 10 $\micron$ would produce obvious 4.07 $\micron$ features in our data, despite the continuum shape. \label{fig:so2spec}}
\end{figure}

While the 4.07 $\micron$ feature of SO$_2$ frost is typically much deeper than the corresponding 3.78 $\micron$ feature, these ratios can change significantly with increasing grain size, as the 4.07 $\micron$ feature broadens and saturates. To investigate whether it is possible to simultaneously explain the 3.78 $\micron$ feature and the 4 $\micron$ region in the Europa spectrum, we used a simple one-component Hapke model \citep{Hapke1981} to produce simulated spectra of SO$_2$ frost for multiple effective particle diameters. Using optical constants from \citet{Schmitt1994}, we produced spectra of solid SO$_2$ at 125 K using effective particle sizes of 10, 100, 1000, and 1500 $\micron$. We then performed continuum subtraction and scaled each spectrum to fit the band area of the observed 3.78 $\micron$ absorption. The fit for a 10 $\micron$ effective particle diameter is shown in Figure \ref{fig:378fits}. While SO$_2$ does not appear to fit the 3.78 $\micron$ feature well, we used the resultant mixing ratios to investigate whether the corresponding 4.07 $\micron$ features would be visible in our spectrum. However, this comparison is complicated by the fact that our data do not span the entire width of the broad 4.07 $\micron$ feature, which extends beyond the edge of the L-band atmospheric transmission window. Because we are lacking data on both sides of the potential feature, we were unable to reliably remove the continuum from our data without making assumptions about its shape. Instead, we linearly mixed the simulated SO$_2$ spectra with our Europa spectrum in the ratios necessary to explain the observed 3.78 $\micron$ feature and assessed the resultant 4 $\micron$ regions by eye. The result for a 10 $\micron$ effective particle size is shown in Figure \ref{fig:so2spec}. 

\begin{figure}
\figurenum{5}
\plotone{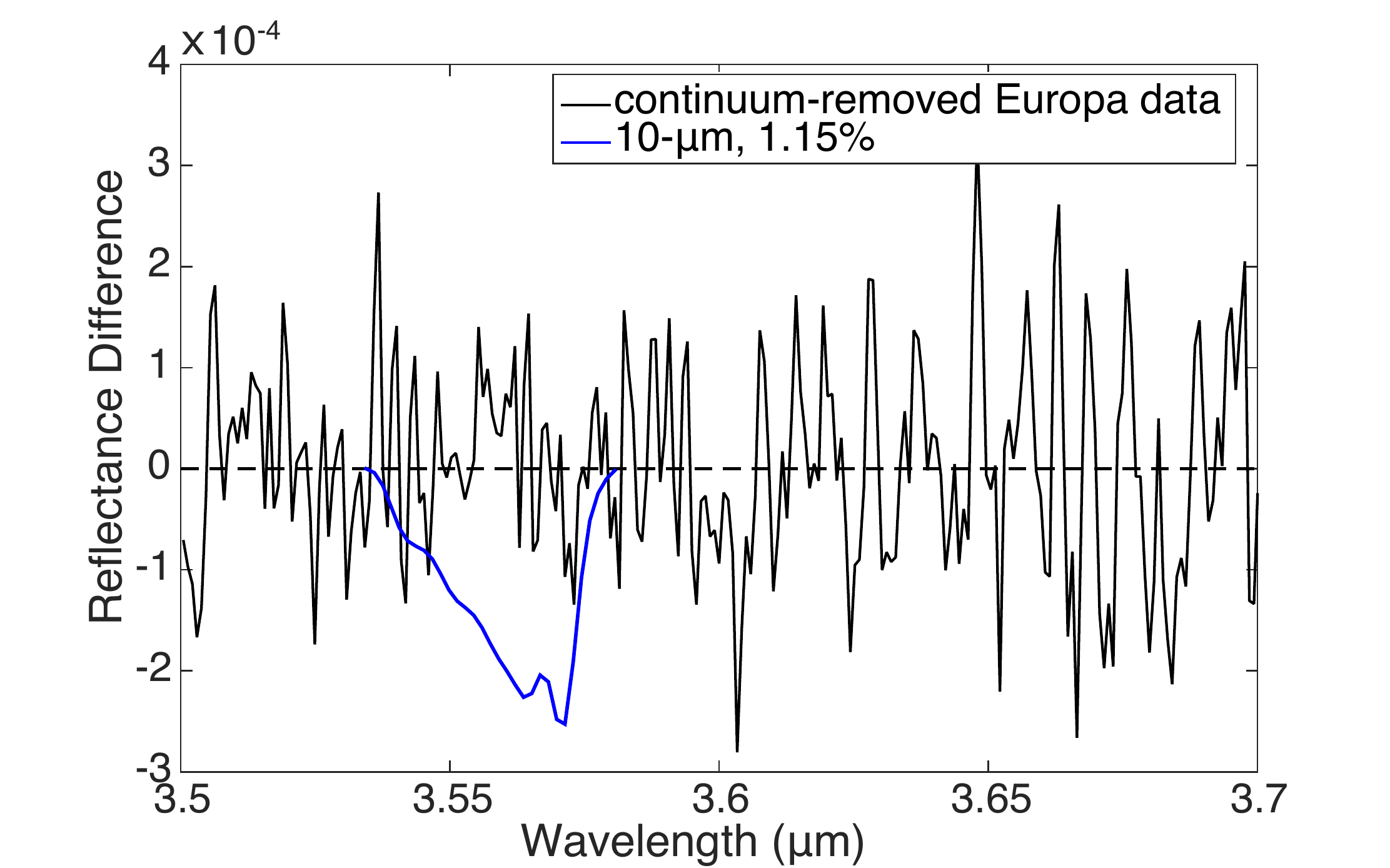}
\caption{Comparison of our continuum-removed Europa spectrum and the continuum-removed 3.56 $\micron$ SO$_2$ feature for an effective grain size of 10 $\micron$. Again, we use the 2011 data of \citet{HandBrown2013} due to its superior signal-to-noise. The SO$_2$ absorption has been scaled by the 1.15\% needed to fit the observed area of the 3.78 $\micron$ feature. Despite the small grain size and relative weakness of this feature, the absorption would be detected in our data, if it were present. \label{fig:4fits}}
\end{figure}

At 10 $\micron$, the effect of the 4.07 $\micron$ feature is clearly distinguishable from the continuum and would be apparent in our data. At sufficiently large grain sizes, however, the 4.07 $\micron$ feature becomes difficult to distinguish from the already dipping Europa continuum in these mixtures. Thus, we do not rule out large SO$_2$ grain sizes on the basis of the 4 $\micron$ region alone. Instead, we use the comparatively small 3.56 $\micron$ absorption of SO$_2$ as a further constraint. For this analysis, we again use the 2011 data of \citet{HandBrown2013} due to its superior quality. In the wavelength region of this feature, the Europa continuum is smooth with a high signal-to-noise, allowing for reliable continuum subtraction. Clipping the wavelength region corresponding to the 3.56 $\micron$ absorption from our data, we fit a fourth-order polynomial continuum and removed it from our Europa spectrum. Then, scaling the continuum-removed 3.56 $\micron$ SO$_2$ feature by the same factor needed to fit the 3.78 $\micron$ feature observed on Europa, we compare the band strengths with our subtracted data (Figure \ref{fig:4fits}).

 Figure \ref{fig:4fits} suggests that SO$_2$ should be detectable at 3.56 $\micron$ in our data, even for a small effective grain size of 10 $\micron$. Indeed, we can put a 5$\sigma$ upper limit of a 0.46\% mixing ratio, which is two and a half times lower than the 1.15\% expected from the strength of the 3.78 $\micron$ feature. While the 4.07 $\micron$ SO$_2$ feature becomes difficult to detect in our data at larger grain sizes, the depth of the 3.56 $\micron$ feature increases relative to that of the 3.78 $\micron$ feature, becoming even more detectable in our data. Conversely, the 4.07 $\micron$ feature becomes stronger relative to the 3.78 $\micron$ feature at grain sizes below 10 $\micron$. Thus, we are confident that the absence of any detected SO$_2$ features at 3.56 or 4.07 $\micron$ in our data excludes the possibility that the observed 3.78 $\micron$ feature is explained by the presence of SO$_2$ on Europa's surface. SO$_2$ is a relatively minor product of the radiolytic sulfur cycle on Europa, the two most abundant and stable being H$_2$SO$_4$ and sulfur chain polymers \citep{Carlson2002}. Thus, while SO$_2$ is likely present on the surface of Europa, it is not in high enough abundance to explain the 3.78 $\micron$ feature. 

\section{Search for Alternative Candidates}\label{sec:alternatives} 
No other confirmed species on the surface of Europa has a 3.78 $\micron$ absorption. Therefore, this feature indicates an undiscovered surface constituent. We searched the USGS, ASTER, and NASA Goddard Cosmic Ice spectral libraries, as well as the \citet{Nyquist1997} compilation of IR spectra, for substances with 3.78 $\micron$ absorptions. Our search was not only constrained to likely radiolytic products, namely sulfur compounds, but it encompassed several categories of interest for Europa, including sulfates, carbonates, nitrates, phosphates, oxides, and various ices relevant to the study of outer solar system bodies. Of all of the species we looked at, only carbonic acid (H$_2$CO$_3$), bloedite (Na$_2$Mg(SO$_4$)$_2$ $\cdot$ 4H$_2$O),  and anhydrite (CaSO$_4$) had absorptions near 3.78 $\micron$ and lacked strong absorptions at other wavelengths that would have been seen by either NIMS or past ground-based work.

\begin{figure}
\figurenum{6}
\plotone{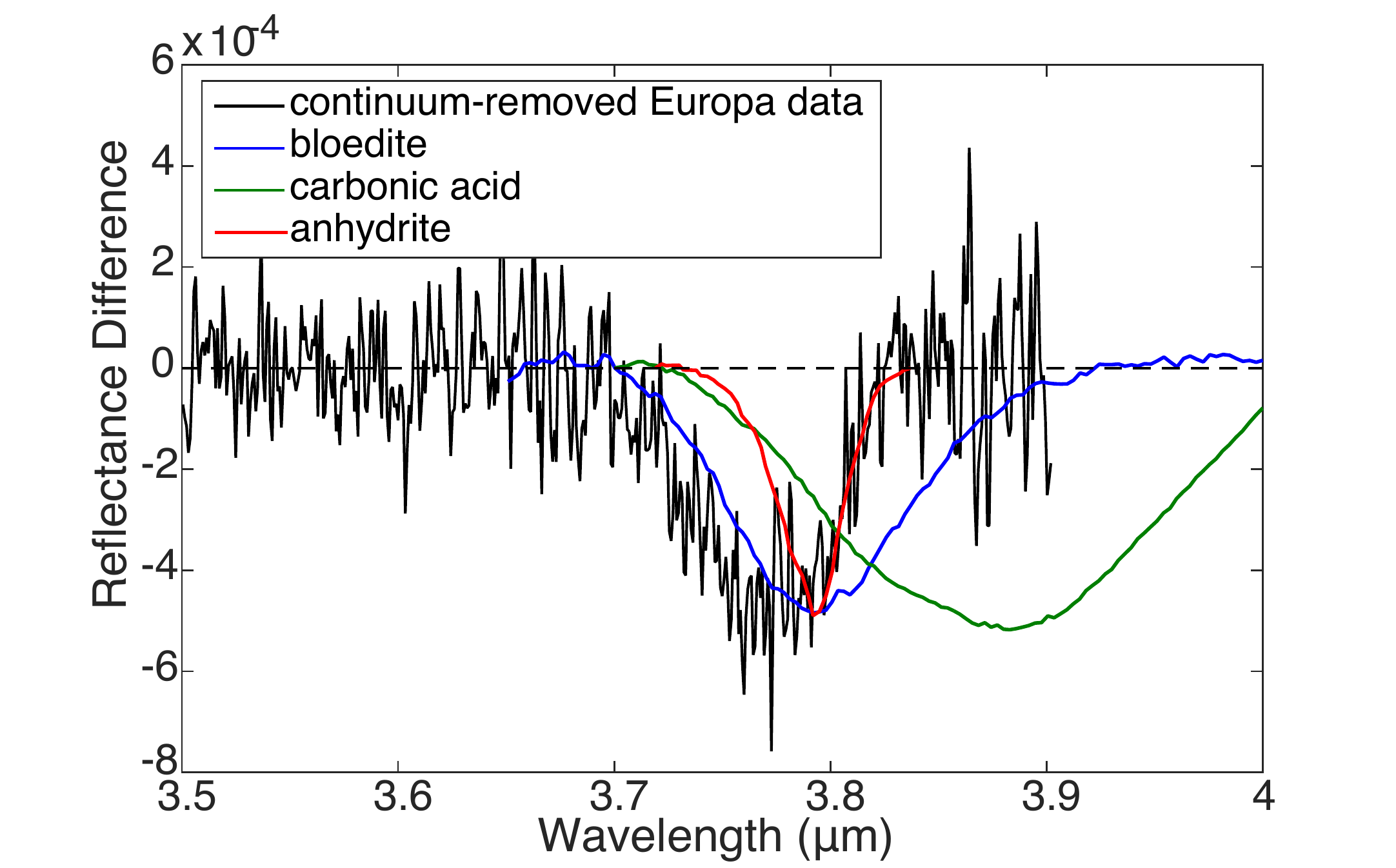}
\caption{Bloedite, carbonic acid, and anhydrite compared to the 3.78 $\micron$ feature of Europa (using the 2011 data of \citet{HandBrown2013}). The bloedite spectrum is taken from the USGS spectral library \citep{USGSlib}, the anhydrite spectrum is from the ASTER spectral library \citep{ASTERlib}, and the carbonic acid feature is radiolytically produced from a mixture of CO$_2$ and water ice at 90 K \citep{Hand2007}. All features were scaled to match the approximate depth of the observed absorption. \label{fig:alternates}}
\end{figure}

Carbonic acid is a radiolytic product that results from the bombardment of CO$_2$ and water ice mixtures with energetic electrons \citep[e.g.][]{Brucato1997, Gerakines2000, Hand2007}, a process that occurs on the trailing hemisphere of Europa. However, laboratory experiments conducted under Europa-like conditions \citep{Hand2007} result in a spectral feature that fits our observed feature poorly (Figure \ref{fig:alternates}). The carbonic acid feature is broader and shifted by $\sim$0.1 $\micron$ toward long wavelengths relative to our detected feature. 

Bloedite, a hydrated sodium-magnesium sulfate, has been previously proposed as one of many potential fits to the 1--2.5 $\micron$ continuum of Europa's surface \citep[e.g.][]{Dalton2005, Carlson2009}. Its presence, as well as the presence of anhydrite (also a sulfate), would seem to fit with the long-held hypothesis that the ocean is dominated by sulfate salts \citep[e.g.][]{Kargel1991, Kargel2000, Fanale2001}. However, the confinement of the 3.78 $\micron$ feature to the trailing hemisphere would suggest that these sulfates participate in the radiolytic chemistry. One possibility is that they begin as chlorides and convert to sulfates upon sulfur ion bombardment, as suggested in \citet{BrownHand2013}. However, neither bloedite nor anhydrite provide good fits to the 3.78 $\micron$ feature. Our observed feature is narrower than that of bloedite and broader than that of anhydrite. In addition, the features of both bloedite and anhydrite are shifted longward of our observed band center. Both library spectra used, however, were taken at Earth temperatures, so it is possible that the fit quality would change under Europa-like conditions.

\section{Conclusions}\label{sec:conclusions}
Using ground-based L-band spectroscopy, we have detected a new absorption feature on the trailing hemisphere of Europa at 3.78 $\micron$. The geography of this spectral feature is consistent with that of a radiation product; however it cannot be explained by any known radiolytic species on the surface. SO$_2$ provides the closest match of known species, but possesses a weak absorption at 3.56 $\micron$ and a strong absorption at 4.07 $\micron$, both of which are absent from our data. Thus, we conclude that the 3.78 $\micron$ feature represents a new, unidentified surface component. Extensive library searches yielded only three plausible candidates---carbonic acid, bloedite, and anhydrite. With the exception of carbonic acid, the majority of available spectra were taken under Earth conditions, and all provided imperfect fits to the data. Ultimately, we are unable to positively identify the source of the 3.78 $\micron$ feature, although there are possible candidates. This highlights the need for more laboratory spectra taken under Europa-like conditions at these wavelengths, particularly of species likely to exist in a different state or mixture than they do on Earth. 

\acknowledgements
This research was supported by Grant 1313461 from the National Science Foundation. K. P. H. acknowledges support from the Jet Propulsion Laboratory, California Institute of Technology, under a contract with the National Aeronautics and Space Administration and funded in part through the internal Research and Technology Development program. The data presented herein were obtained at the W. M. Keck Observatory, which is operated as a scientific partnership among the California Institute of Technology, the University of California, and the National Aeronautics and Space Administration. The Observatory was made possible by the generous financial support of the W. M. Keck Foundation. The authors wish to recognize and acknowledge the very significant cultural role and reverence that the summit of Mauna Kea has always had within the indigenous Hawaiian community. We are most fortunate to have the opportunity to conduct observations from this mountain. The authors thank Bethany L. Ehlmann, George R. Rossman, and Robert P. Hodyss for helpful discussions. 

\software{Astropy \citep{Astropy}, skimage.transform \citep{skimage}}

\end{document}